\documentstyle[12pt]{article}

\begin{document}
\newcommand{\dis}{\displaystyle}
\newcommand{\id}{ 1 \hspace{-2.85pt} {\rm I} \hspace{2.5mm}}
{\thispagestyle{empty}
\rightline{} 
\rightline{} 
\rightline{} 
\vskip 1cm
\centerline{\large \bf A Labelling Scheme for}
\centerline{\large \bf Higher Dimensional Simplex Equations}


\vskip 1cm
\centerline{ L. C. Kwek {\footnote{E-mail address:
SCIP3057@LEONIS.NUS.SG }} and C. H. Oh {\footnote{E-mail address:
PHYOHCH@NUS.SG }} }
\centerline{{\it Department of Physics, Faculty of Science, } }
\centerline{{\it National University of Singapore,Lower Kent Ridge,} }
\centerline{{\it Singapore 0511, Republic of Singapore. } }
\vskip 0.1in

\vskip 1cm
\centerline{\bf Abstract} \vspace{10mm}

\noindent We present a succinct way of obtaining all
possible higher dimensional generalization of Quantum Yang-Baxter
Equation (QYBE).  Using the scheme, we could generate the two popular
three-simplex equations, namely: Zamolodchikov's
tetrahedron equation (ZTE) and Frenkel and Moore equation (FME).

\newpage

The Quantum Yang-Baxter Equation (QYBE) is a non-linear equation which
appears in various forms in areas like integrable statistical models,
topological field theories, the theory of braid groups,the theory of
knots and links and conformal field theory.

Several higher dimensional generalizations of QYBE have been proposed.
By considering the scattering of straight strings in $2 + 1$
dimensions, Zamolodchikov proposed a higher dimensional generalization
of QYBE, commonly called the tetrahedron equation (ZTE) \cite{Zamo}:
\begin{equation}
R_{123}R_{145}R_{246}R_{356} = R_{356}R_{246}R_{145}R_{123}, \label{tetra}
\end{equation}
where $R_{123} = R \otimes \id$ etc and $R \in \ {\rm End}(V \otimes V
\otimes V)$ for some vector space $V$. As the QYBE is often called a
two-simplex equation, this generalization of QYBE is a three-simplex
equation. In general, there are $d^{12}$
equations with $d^6$ variables, where $d$ is the dimension of the
vector space $V$.    Despite its complexity, Zamolodchikov has ingeniously
proposed a spectral dependent solution which is subsequently confirmed
by Baxter \cite{Zamo, Baxter}.

\medskip

The tetrahedron equation is not the only
possible higher dimensional generalization.  Frenkel and Moore
\cite{Frenkel} has proposed another higher dimensional generalization (FME),
namely

\begin{equation}
R_{123}R_{124}R_{134}R_{234} = R_{234}R_{134}R_{124}R_{123}, \label{fme}
\end{equation}

\noindent They have also given an analytical solution for their three-simplex
equation, namely:
\begin{equation}
R = q^{1/2}q^{(h \otimes h \otimes h)/2} [1 + (q - q^{-1})
(h \otimes e \otimes f + f \otimes e \otimes h
- e \otimes h q^{-h} \otimes f)], \label{fmesoln}
\end{equation}
where $e$, $f$, $h$ are generators of $sl(2)$ satisfying $[h,e] = 2e$, $[h,f]
= 2f$, $[e,f]= h$.
The three-simplex equation is not equivalent to Zamolodchikov's
equation as explained by Frenkel and Moore in their paper
\cite{Frenkel}.
Furthermore, solutions of the three-simplex equation do not in general
satisfy Zamolodchikov's tetrahedron equation (\ref{tetra}). In fact,
the expression (\ref{fmesoln})  does not satisfy Zamolodchikov's
tetrahedron equation unless the parameter q approaches unity, giving
the identity matrix.

By considering the commutativity of the matrices  $ S_{i_{1}
i_{2} \cdots i_{d}}^{j_{1} j_{2} \cdots j_{d}}$, Maillet and Nijhoff
\cite{maillet} was able to generalized the QYBE to higher dimensional
forms.

In this letter, we present a succinct way of writing these
generalized equations. Consider a sequence $(1, 1, 0, \cdots, 0)$, in
which all the entries are zeros with only two ones, and all its
possible permutations. By interpreting each sequence as a binary number
and arranging the resulting array in decreasing ( or increasing) value,
we could obtain a new array in which the `transpose' generates the
higher dimensional forms of the QYBE.

As an example, consider the sequence $(1,1,0,0)$.  The resulting array
obtained by arranging all possible permutations into an array of
decreasing binary values is:

\begin{equation}
\left[ \begin{array}{cccc}
1 & 1 & 0 & 0 \\
1 & 0 & 1 & 0 \\
1 & 0 & 0 & 1 \\
0 & 1 & 1 & 0 \\
0 & 1 & 0 & 1 \\
0 & 0 & 1 & 1  \label{frtarr}
\end{array}
\right]
\end{equation}

\noindent{By writing rows as columns and columns as rows, we get a new
array:}

\begin{equation}
\left[ \begin{array}{cccccc}
1 & 1 & 1 & 0 & 0 & 0 \\
1 & 0 & 0 & 1 & 1 & 0 \\
0 & 1 & 0 & 1 & 0 & 1 \\
0 & 0 & 1 & 0 & 1 & 1
\end{array} \right] \label{newarray}
\end{equation}

\noindent Identifying the first row of the new matrix (\ref{newarray}) as the
transfer matrix $R_{123}$ acting on the vector spaces $V \otimes V
\otimes V \otimes \id \otimes \id \otimes \id $ and so forth, we obtain
$R_{123}R_{145}R_{246}R_{356} $, which is the LHS of equation
(\ref{tetra}).

Consider the `mirror' image of (\ref{frtarr}), we get the array:

\begin{equation}
\left[ \begin{array}{cccc}
0 & 0 & 1 & 1 \\
0 & 1 & 0 & 1 \\
1 & 0 & 0 & 1 \\
0 & 1 & 1 & 0 \\
1 & 0 & 1 & 0 \\
1 & 1 & 0 & 0 \\
\end{array}
\right]
\end{equation}

\noindent and consider the transpose again, we get:

\begin{equation}
\left[ \begin{array}{cccccc}
0 & 0 & 1 & 0 & 1 & 1 \\
0 & 1 & 0 & 1 & 0 & 1 \\
1 & 0 & 0 & 1 & 1 & 0 \\
1 & 1 & 1 & 0 & 0 & 0
\end{array} \right] \label{newarray2}
\end{equation}

The resulting array is the same as  the array
(\ref{newarray}) but with the row order reversed.
The first row is now identified as $R_{356}$ and so forth, and the
RHS of equation(\ref{tetra}) is obtained.

We note that if we start with the starting sequence $(1, 1, 0)$, we get
QYBE.  Further, if we consider the sequence $(1, 1, 0, 0, 0)$, we get
Bazhanov-Stroganov equations \cite{maillet}, namely:

\begin{eqnarray}
 & &R_{1,2,3,4} R_{1,5,6,7}R_{2,5,8,9} R_{3,6,8,10}
R_{4,7,9,10}  \nonumber \\
& = & R_{4,7,9,10} R_{3,6,8,10}  R_{2,5,8,9}
R_{1,5,6,7}  R_{1,2,3,4}.
\end{eqnarray}

\noindent Higher dimensional forms of the
tetrahedron equation (\ref{tetra}) can be obtained by considering
sequences of the form $(1, 1, 0, \cdots, 0)$.

The Frenkel and Moore equation (\ref{fme}) and its
higher dimensional generalizations can be generated from starting sequences
of the form $(1, 1,\cdots, 1,0)$, in which all the entries except
the last one are ones.  These sequences give trivial permutation
arrays.  In particular, the sequence $(1,1,1,0)$ gives the Frenkel
and Moore equation (\ref{fme}).

When the length of the starting sequence exceeds four, new simplex
equations are obtained. For example, the sequence $(1, 1, 1, 0, 0)$
yields a new commutativity equation:

\begin{eqnarray}
 & &R_{1,2,3,4,5,6} R_{1,2,3,7,8,9} R_{1,4,5,7,8,10} R_{2,4,6,7,9,10}
R_{3,5,6,8,9,10}  \nonumber \\
& = & R_{3,5,6,8,9,10} R_{2,4,6,7,9,10}  R_{1,4,5,7,8,10}
R_{1,2,3,7,8,9}  R_{1,2,3,4,5,6}.
\end{eqnarray}

\noindent As we increase
the length of the sequence, we find an
increasing number of such higher dimensional forms.

An alternative way of looking at the labelling scheme is to consider
the non-commutative expansion:

\begin{equation}
(x_1 + y_1)(x_2 + y_2) \cdots (x_n + y_n) = (y_1 + x_1)(y_2 + x_2)
\cdots (y_n + x_n)
\end{equation}

\noindent where we identify for instance, in the case $n = 4$, the term $x_1
x_2 y_3 y_4$ as
the sequence $(1,1,0,0)$, setting $x_i = 1$ and $y_i = 0$.

As an example, if we consider the expansion:

\begin{equation}
(x_1 + y_1)(x_2 + y_2)(x_3 + y_3)  =  (x_3 + y_3)(x_2 + y_2)(x_1 +
y_1) , \label{ybe}
\end{equation}

\noindent the LHS of the expansion gives

\begin{equation}
x_1 x_2 x_3 + x_1 x_2 y_3 + x_1 y_2 x_3 + x_1 y_2 y_3 + y_1 x_2 x_3 +
y_1 x_2 y_3 + y_1 y_2 x_3 + y_1 y_2 y_3 .
\end{equation}

\noindent Replacing $x_i$ by $1$ and $y_i$ by $0$ and grouping the
terms with the same number of $x$'s into arrays of the form
(\ref{frtarr}), we get four different arrays, namely:

\begin{equation}
\left[ \begin{array}{ccc}
1 & 1 & 1 \label{yba}
\end{array}
\right]
\end{equation}
\noindent corresponding to $x_1 x_2 x_3$,

\begin{equation}
\left[ \begin{array}{ccc}
1 & 1 & 0 \\
1 & 0 & 1 \\
0 & 1 & 1 \label{ybb}
\end{array}
\right]
\end{equation}

\noindent corresponding to terms of form $x_i x_j y_k $,

\begin{equation}
\left[ \begin{array}{ccc}
1 & 0 & 0 \\
0 & 1 & 0 \\
0 & 0 & 1 \label{ybc}
\end{array}
\right]
\end{equation}

\noindent corresponding to terms of the form $x_i y_j y_k$, and

\begin{equation}
\left[ \begin{array}{ccc}
0 & 0 & 0 \label{ybd}
\end{array}
\right]
\end{equation}

\noindent corresponding to terms of the form $y_i y_j y_k$.

The first array (\ref{yba}), the third array (\ref{ybc}) and the fourth
array (\ref{ybd}) lead to trivial simplex equations, namely  $R_{1234} =
R_{1234} $, $R_1 R_2 R_3
R_4 = R_4 R_3 R_2 R_1 $ and $\id = \id$.  Only terms of the array
(\ref{ybb}) provide us with non-trivial simplex equations, namely
$R_{12}R_{13}R_{23}$, the LHS of the QYBE. Expanding the RHS of eq
(\ref{ybe}) and grouping the terms $x_i x_j y_k $ will lead to
$R_{23}R_{13}R_{12}$, the RHS of the QYBE.

The number of terms in each array corresponds to the coefficient of the
binomial expansion $(x + y)^n$, in our previous example, we have
corresponding to each array:
$$\dis {3 \choose 3}, {3 \choose 2}, {3
\choose 1}, \mbox{~ and ~} {3 \choose 0}. $$ Thus, it is  not
surprising that the number of vector spaces in which the higher
dimensional generalization of the QYBE acts are the coefficients of a
binomial expansion. Further, if we consider the expansion $(x_1 +
y_1)(x_2 + y_2)(x_3 + y_3)(x_4 + y_4)$, we see that for three-simplex
case, there are only two possible non-trivial generalizations, namely
ZTE and FME. Work is still in progress to understand more about the
implications and significance of this labelling scheme, here we merely
note that Yang-Baxter equation is deeply rooted to the permutation
group \cite{yang}.


\newpage

\begin{thebibliography}{99}

\bibitem{Zamo} A.B. Zamolodchikov 1981 Commun. Math. Phys.  {\bf 79}
489
\bibitem{Baxter} R.J. Baxter 1986 Physica {\bf 18 D} 321
\bibitem{Frenkel} I. Frenkel and G. Moore 1991 Commun. Math. Phys.
{\bf 138} 259
\bibitem{maillet} J. M. Maillet and F. Nijhoff 1990 In {\it Non-Linear
Evolution
Equations: Integrability and Spectral Methods} 537 edited by A.
Degasperis, A.P. Fordy and M. Lakshmanan (Manchester University
Press)
\bibitem{yang} C. N. Yang 1991 In {\it The Oskar Klein Memorial Lectures,
Vol 1: Lectures by C N Yang and S Weinberg} 35 edited by G. Eksporg
(World Scientific: Singapore)

\end{thebibliography}
\end{document}